\documentclass[a4paper,11pt]{article}
\usepackage{pos}

\title{What shall we learn from a future supernova?}

\author*[a]{M. Cristina Volpe}

\affiliation[a]{CNRS, Université Paris Cit\'e, Astroparticule et Cosmologie, F-75013 Paris, France}

\emailAdd{volpe@apc.univ-paris7.fr}

\abstract{Core-collapse supernovae constitute a unique laboratory for particle physics and astrophysics. They are powerful neutrino sources of all flavors, emitting essentially all the gravitational binding energy through neutrinos, at the end of their life. I will highlight how crucial is the observation of the next core-collapse supernova and of the diffuse supernova neutrino background, whose discovery might be imminent.}

\FullConference{12th Neutrino Oscillation Workshop (NOW2024)\\
 2-8, September 2024\\
Otranto, Lecce, Italy\\}

\begin{document}
\maketitle

\section{Introduction}
\noindent
In the last thousand years, only six supernova events were observed in our Galaxy. Three of them were SNIa, namely SN1006, SN1572, SN1604. The other three were core-collapse supernovae, i.e.
SN1054, SN1181 and SN1667 (Cas A) \cite{Bethe:1990mw,Adams:2013ana}. The Milky Way is one of the six largest galaxies of the Local Group that include the Large and the Small Magellanic Clouds, 
the Triangulum Galaxy, NGC3109 and Andromeda (M31). In the last century two further massive stars were observed in the Local Group, namely SN1885 in Andromeda and the famous SN1987A 
for which M. Koshiba received the 2002 Physics Nobel Prize (1/4) with R. Davis (1/4) for the pioneering observation of solar neutrinos and R. Giacconi (1/2) for X-ray astronomy.
 
Core-collapse supernovae\footnote{Supernovae are classified based on their spectral properties. In particular, supernovae of type Ia have hydrogen in their spectra but no silicium, and end their life as thermonuclear explosions. Core-collapse supernovae are of type II if their spectra include H and He; whereas they are of type  Ib and Ic if their spectra have no H, but are rich or poor in He, respectively.} spend most of their life burning H (about 7 Mya for a 25 $M_{\rm Sun}$ star) and He (about 5 $10^5$ yrs) whereas the last stages of C and Si burning last only  about 600 days and a few days, before the core collapses.  Then, about $3 \times 10^{53}$ ergs are taken away by neutrinos of all flavors in a burst during ten seconds, leaving as supernova remnant either a neutron-star, or a black-hole. 
 
 Core-collapse supernovae are rare events in our Milky Way.  Analyses of their rate per century yields $1.9 \pm 1.1$ events based on $^{26}$Al in our Galaxy \cite{Diehl:2006cf}, $3.2^{+7.3} _{-2.6}$ events from historical supernovae in the Milky Way \cite{Adams:2013ana}, observations of supernovae\footnote{This analysis considers supernovae in different morphological classes of host galaxies.} or Galactic neutron-stars provide $1.7 \pm 0.74$ events \cite{Cappellaro:1993ns} and $7.2 \pm 2.7 $ events\footnote{This analysis gives a high supernova rate compared to others, and might have a problem of counting \cite{Keane:2008jj}.} \cite{Keane:2008jj} respectively, the counting of massive stars at 1.5 kpc nearby the Sun indicate 1-2 events \cite{Reed:2005en}. Finally the combination of several of these estimates gives the rate of $1.63 \pm 0.46$/century \cite{Rozwadowska:2020nab}.  
 Clearly observing the next supernova would be a lucky and crucial event that will bring a harvest of information both for astrophysics and for particle physics. 

Core-collapse supernovae present many similarities and some differences with other dense environments (binary neutron-star mergers, accretion disks around black holes and the early Universe). According to our current understanding, flavor mechanisms emerge due to the neutrino-neutrino interaction which is sizable in dense environments, collisions, shock wave effects and turbulence. While significant progress has been made, important questions remain open, e.g. the role of correlations beyond the mean-field and the effects of strong gravitational fields nearby compact objects (neutron stars or black holes). These developments have an intrinsic theoretical interest and are also essential for future observations of the next core-collapse supernova, of the diffuse supernova background, for the explosion mechanism and for identifying the site(s) where the $r$-process takes place (see e.g. \cite{Volpe:2023met} for a review).
 
Importantly neutrinos from past supernovae that constitute the relic, or diffuse, supernova neutrino background (DSNB) might be imminent. The DSNB encodes information from particle physics, astrophysics and cosmology, and is thus complementary to single supernovae (see \cite{Volpe:2023met,Beacom:2010kk,Mathews:2014qba,Ando:2023fcc} for reviews). In particular, the DSNB, like supernovae, has a unique sensitivity to neutrino non-radiative decay (see \cite{Ivanez-Ballesteros:2022szu} and references therein). 

The combined analysis of SK-I to SK-IV data (twenty years of data taking) in Super-Kamiokande found a 1.5 $\sigma$ excess over the background prediction (model dependent analysis). 
Interestingly, the combined sensitivity of SK-I to SK-IV data is on par with 4 DSNB predictions that have very different inputs  \cite{Super-Kamiokande:2021jaq}.
Moreover, the analysis of the first results of Super-Kamiokande data with Gadolinium addition (SK-VI to SK-VII) increased the statistical significance to 2.3 $\sigma$ \cite{Harada:2024}.
If the excess found is indeed a signal, then the upcoming Hyper-Kamiokande, JUNO and DUNE will have the sensitivity 
for the DSNB discovery.
This will open a novel and unique observational window in low energy neutrino astrophysics.
 
 \section{What shall we learn from a future supernova?}
 \noindent
 The question is vast and was addressed by several reviews (see e.g. \cite{Volpe:2023met,Horiuchi:2018ofe,Mirizzi:2015eza}).
Here I will highlight some of the key aspects for which future supernova observations will be crucial, or unique.
 
 The few MeVs neutrinos emitted at the late stages before the core collapses encode information on the progenitor and on the late stages of the life of massive stars. Thus, their observation would confirm stellar evolution theory \cite{Odrzywolek:2003vn,Patton:2017neq,Kato:2020hlc,Kato:2020lwd,Ando:2023fcc}.  A list of potential red-supergiant candidates in the detectable range for pre-supernova neutrinos (less than 1 kpc) is shown in Table 4 of Ref. \cite{Kato:2020hlc}. 
 These pre-supernova neutrinos could be detected if the supernova is close enough, e.g. by KamLAND, at $3 \sigma $, for a 25 $M_{\rm Sun}$ star at 690 pc, 2 days before the explosion \cite{KamLAND:2015dbn}; whereas the pre-supernova alert system of KamLAND combined with Super-Kamiokande would give an alert for a 15 $M_{Sun}$ star within 510 pc (e.g. about 12 hr in advance for Betegeulse) \cite{KamLAND:2024uia}. Moreover the detection of the neutrino emission after 10 s would tell us about the fate of the supernova, the equation of state and the cooling of the newly born neutron star \cite{Li:2020ujl}. 
 
The observation of the pre-supernova and/or the supernova neutrinos would give an early alert to the international community and also to amateurs astronomers thanks also to the Supernova Early Warning System. In the current version (2.0), the network of neutrino detectors to observe the next supernova early designed to be prompt, positive and pointing, is now including dark matter detectors, and gravitational wave detectors for multi-messenger astronomy \cite{SNEWS:2020tbu}.  

On February 23rd, 1987, the blue supergiant Sk-69$^{\circ}$202 exploded at 50 kpc from the Earth, giving SN1987A. To date, it is the only supernova for which we observed neutrinos from the core. The measurement of 24 $\bar{\nu}_e$ events brought confirmation of the expected neutrino time signal, average energy and total luminosity\footnote{Under the equipartition hypothesis among the neutrino species.}. Moreover the Bayesian analysis of the neutrino fluences performed by Ref. \cite{Loredo:2001rx} rejected the prompt model of Colgate and White and confirmed the delayed neutrino-heating mechanism by Wilson and Bethe and Wilson \cite{Bethe:1985sox}. These important findings were also confirmed by the analysis of Ref. \cite{Pagliaroli:2008ur}.
After more than thirty years searches, SN1987A remnant was identified: a dust-obscured thermally emitting neutron star \cite{Alp:2018oek, Cigan:2019shp,Page:2020gsx}.  

SN1987A observations represent a unique laboratory for astrophysics and particle physics and is a telling example of how much we can learn from such rare events. Analyses of SN1987A observations and of the neutrino events keep providing constraints on neutrino unknown properties, new particles and interactions. For example, based on a 3$\nu$ framework, a 7D-likelihood analysis of the 24 neutrino 
events measured by Kamiokande, IMB and Baksan provided the limit for neutrino non-radiative two-body decay of 
$\tau/m > 2.4~(1.2)~\times~10^5$ s/eV at 68$\% ~(90~ \%$) C.L. (lifetime-to-mass ratio for the mass eigenstates $\nu_2$ and $\nu_1$, inverted mass ordering). This lower bound is tighter than those obtained by Earth-based experiments and is competitive with the model-dependent limits from cosmology.                                                                                   

If the next galactic supernova is located at 10 kpc, we will be able to measure from several hundred events in detectors like HALO-2 or KamLAND, 3 $10^3$ and 8 $10^3$ events in DUNE and JUNO respectively, $10^4$ in Super-Kamiokande, 10$^5$ in Hyper-Kamiokande, and 10$^6$ in IceCUBE. On the other hand,  we will measure events associated with coherent neutrino-nucleus scattering in dark matter detectors, for example 120 events are expected in Xenon nT\footnote{Note that the Xenon nT Collaboration just observed the first events from solar $^8$B neutrinos \cite{XENON:2024ijk}.}, about 700 events in DARWIN and 336 events in DarkSide-20 kt (see \cite{Volpe:2023met} and references therein).  

Understanding the explosion mechanism of massive stars is a six-decade quest. Currently there is an emerging consensus that the majority of supernovae explodes due to the 
delayed neutrino-heating mechanism where neutrinos efficiently reheat the shock aided by neutrino-driven convection, 
turbulence and hydrodynamic instabilities, in particular the standing accretion shock instability (SASI). This paradigm was built through an international effort which has produced a major step
forward every decade \cite{Mezzacappa:2020pkk}. Successful explosions are obtained in 2D and also 3D models for different progenitors, although currently there is not yet a consensus on the role of SASI for all progenitors.  

Clearly, the detection of next (extra)galactic supernova will be crucial to confirm the current paradigm for the explosion mechanism. In particular, studies have shown that the SASI will 
leave characteristic direction-dependent imprints in the neutrino time signal \cite{Tamborra:2014hga,Muller:2014rpb,Walk:2019miz}. As for the total neutrino luminosity, a 10-dimensional likelihood analysis of inverse-beta decay events in Super-Kamiokande and Hyper-Kamiokande 
showed it could be determined with a precision of 11 $\%$ and 3 $\%$ respectively, assuming only the Mikheev-Smirnov-Wolfenstein (MSW) effect takes place, even if one does not fix any of the neutrino flux parameters. From this information one could for example retrieve the compactness of the newly born neutron star, through the knowledge of the neutron-star equation of state \cite{GalloRosso:2017hbp} which is progressing much thanks to gravitational wave observations.  

The detection of the $\nu_e$ from the neutronisation peak would yield confirmation of the early phase of the explosion. 
The about 20 ms neutrino emission appears to be influenced by the MSW effect only, and therefore is a good ground to explore non-standard neutrino properties such as neutrino non-radiative decay or non-standard neutrino-matter interactions \cite{deGouvea:2019goq,Das:2017iuj}. Moreover, the neutronisation burst could  be used for the concurrent detection of gravitational waves. Indeed gravitational waves of various frequencies can be produced from the core-bounce, from neutrino-driven convection, neutrino-heating in the gain layer and the SASI. While predictions agree on the overall picture, they disagree on pinning down the specific contributions (see e.g. Ref. \cite{Mezzacappa:2024zph} for a discussion). 

\begin{figure}
\begin{center}
\includegraphics[scale=0.37]{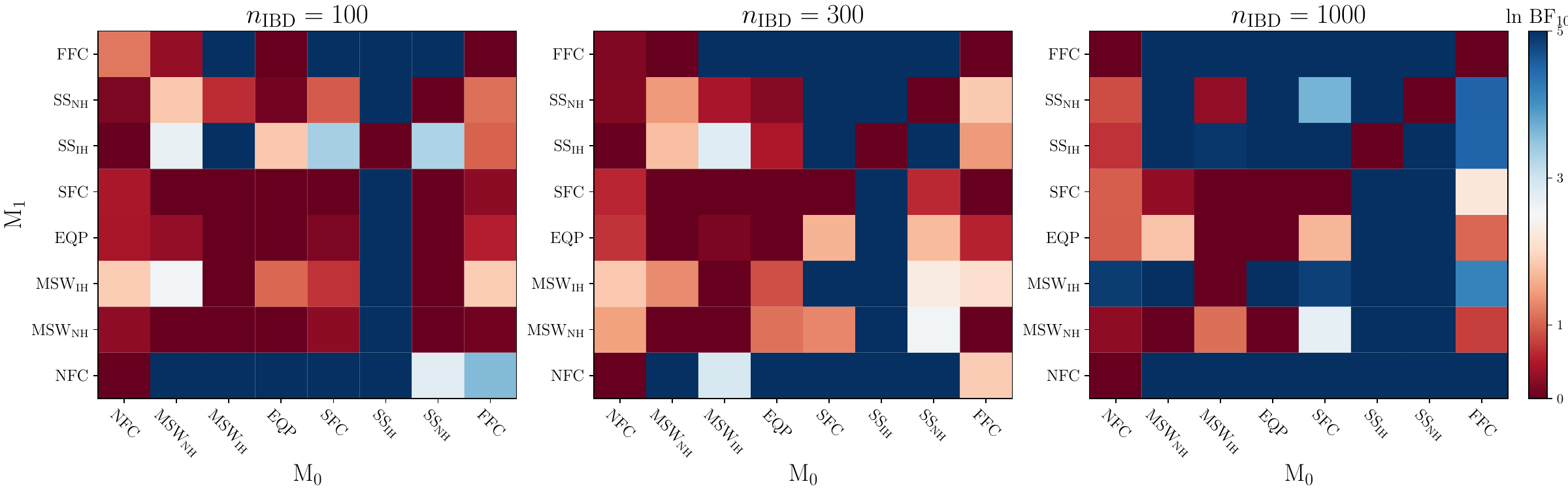}
\caption{Heatmaps for the Bayes factor based on 100 (left), 300 (middle) to 1000 (right figure) inverse-beta decay events from the accretion phase of an exploding supernova. The location of the supernova is unknown and the neutrino flux parameters are not fixed. The acronysms correspond to no-flavor conversion (NFC), the Mikheev-Smirnov-Wolfenstein effect for normal (MSW$_{\rm NH}$) or inverted (MSW$_{\rm IH}$) neutrino mass ordering, flavor equipartition (EQP), slow flavor conversion (SFC) modes, spectral swapping (SS$_{\rm NH}$, SS$_{\rm IH}$) fast flavor conversion (FFC). If the logarithm of the Bayes factor is : i) (0-1), it is not significant; ii) (1–3), it indicates positive evidence; iii) (3–5): it reflects strong evidence; iv) it is > 5, it demonstrates very strong evidence.  (see \cite{Abbar:2024nhz} for details).}
\label{fig:flav}
\end{center}
\end{figure}

Since almost twenty years, an important international effort has been made to unravel how neutrinos evolve in dense environments (see the review \cite{Volpe:2023met,Mirizzi:2015eza}). This is indeed a complex weakly-interacting many-body problem. To determine the neutrino flavor evolution while neutrinos traverse the medium requires the solution of coupled non-linear equations. Until recently such equations were solved in the mean-field approximation, including the tadpole diagrams of the charged- and neutral-current interaction terms of the Glashow-Weinberg-Salam Model. 

Let us mention that the validity of the mean-field approximation was already questioned more than ten years ago. Indeed Refs.\cite{Volpe:2013uxl,Balantekin:2006tg} pointed out that there were corrections at the mean-field level missing, as well as beyond the mean-field. It was also discussed that such corrections could produce effects in the "transition" region, where it was usually argued that the collision terms and the mean-field terms were well separated under an argument of separation of scales. 
This ansatz has finally been overcome in particular with the emergence of very short scale modes called fast \cite{Sawyer:2015dsa}.
Currently there are ongoing efforts toward the solution of the neutrino kinetic equations, started with e.g. Ref. \cite{Capozzi:2018clo} that first studied the interplay between collisions and fast modes, whereas Ref. \cite{Johns:2021qby} identified a new flavor mechanism, termed "collisional instability".

Several studies investigated our ability to identify supernova models through Bayesian analyses. Considering the accretion phase, Ref. \cite{Hyper-Kamiokande:2021frf} have explored the capability to discriminate among five (one- or multidimensional) supernova models from different groups; whereas Ref. \cite{Saez:2024ayk} has studied 18 2D and 3D models from 9 $M_{\rm Sun}$ to 60 $M_{\rm Sun}$ progenitors. On the contrary Ref.\cite{Olsen:2022pkn} has considered the long-term emission of seven one-dimensional models (different progenitor and equation of state). These investigations came to the conclusion that the possibility to disentangle models, if a new supernova is observed, is in many cases excellent. 

While previous studies only included the MSW effect, Ref. \cite{Abbar:2024nhz} 
performed the first Bayesian analysis to address our ability to identify neutrino flavor mechanisms in a supernova.
Figure \ref{fig:flav} presents heatmaps obtained considering a variable number of inverse-beta decay events during the accretion phase, without assuming the parameters defining the neutrino fluxes to be known (but only some reasonable priors). The distance to the supernova is also assumed not to be known. One can see that even 1000 IBD events appear to be sufficient to discriminate among different scenarios of flavor evolution. 

In conclusion, if the next supernova is observed through its neutrinos, the gravitational waves and its electromagnetic emission (thanks for example to the All-Sky Automated Survey for Supernovae \cite{Hart:2023rqs}), we might determine the late stages of a massive star evolution, learn about the supernova progenitor and its localization, determine the mass-radius relation of the newly born neutron star, confirm the explosion mechanism, learn about the 
neutron-star equation of state at finite temperature, search for non-standard neutrino properties such as neutrino non-radiative decay to Majorons \cite{Ivanez-Ballesteros:2024nws}, interactions and new physics. Moreover, the observation of the next supernova will be a unique laboratory to study neutrino flavor conversion in dense environments, a complex many-body problem where an impressive theoretical progress has been made in the last decades. In this respect recent Bayesian analyses appear promising.


\begin{thebibliography}{99}
\bibitem{Bethe:1990mw}
H.~A.~Bethe,
Rev. Mod. Phys. \textbf{62}, 801-866 (1990)

\bibitem{Adams:2013ana}
S.~M.~Adams, C.~S.~Kochanek, J.~F.~Beacom, M.~R.~Vagins and K.~Z.~Stanek,
Astrophys. J. \textbf{778}, 164 (2013)
[arXiv:1306.0559 [astro-ph.HE]].

\bibitem{Diehl:2006cf}
R.~Diehl, H.~Halloin, K.~Kretschmer, G.~G.~Lichti, V.~Schoenfelder, A.~W.~Strong, A.~von Kienlin, W.~Wang, P.~Jean and J.~Knoedlseder, \textit{et al.}
Nature \textbf{439}, 45-47 (2006)
[arXiv:astro-ph/0601015 [astro-ph]].

\bibitem{Cappellaro:1993ns}
E.~Cappellaro, M.~Turatto, Benetti, D.~Y.~Tsvetkov, O.~S.~Bartunov and I.~N.~Makarova,
Astron. Astrophys. \textbf{273}, 383 (1993)
[arXiv:astro-ph/9302017 [astro-ph]].

\bibitem{Keane:2008jj}
E.~F.~Keane and M.~Kramer,
Mon. Not. Roy. Astron. Soc. \textbf{391}, 2009 (2008)
[arXiv:0810.1512 [astro-ph]].

\bibitem{Reed:2005en}
B.~C.~Reed,
Astron. J. \textbf{130}, 1652-1657 (2005)
[arXiv:astro-ph/0506708 [astro-ph]].

\bibitem{Rozwadowska:2020nab}
K.~Rozwadowska, F.~Vissani and E.~Cappellaro,
New Astron. \textbf{83}, 101498 (2021)
[arXiv:2009.03438 [astro-ph.HE]].



\bibitem{Volpe:2023met}
M.~C.~Volpe,
Rev. Mod. Phys. \textbf{96}, no.2, 025004 (2024)
[arXiv:2301.11814 [hep-ph]].

\bibitem{Beacom:2010kk}
J.~F.~Beacom,
Ann. Rev. Nucl. Part. Sci. \textbf{60}, 439-462 (2010)
[arXiv:1004.3311 [astro-ph.HE]].

\bibitem{Mathews:2014qba}
G.~J.~Mathews, J.~Hidaka, T.~Kajino and J.~Suzuki,
Astrophys. J. \textbf{790}, 115 (2014)
[arXiv:1405.0458 [astro-ph.CO]].

\bibitem{Ando:2023fcc}
S.~Ando, N.~Ekanger, S.~Horiuchi and Y.~Koshio,
[arXiv:2306.16076 [astro-ph.HE]].

\bibitem{Ivanez-Ballesteros:2022szu}
P.~Ivanez-Ballesteros and M.~C.~Volpe,
Phys. Rev. D \textbf{107}, no.2, 023017 (2023)
[arXiv:2209.12465 [hep-ph]].

\bibitem{Super-Kamiokande:2021jaq}
K.~Abe \textit{et al.} [Super-Kamiokande],
Phys. Rev. D \textbf{104}, no.12, 122002 (2021)
[arXiv:2109.11174 [astro-ph.HE]].

\bibitem{Harada:2024}
M. Harada,
Talk given at NEUTRINO 2024, Milano, 17th-21st June (2024).

\bibitem{Horiuchi:2018ofe}
S.~Horiuchi and J.~P.~Kneller,
J. Phys. G \textbf{45}, no.4, 043002 (2018)
[arXiv:1709.01515 [astro-ph.HE]].

\bibitem{Mirizzi:2015eza}
A.~Mirizzi, I.~Tamborra, H.~T.~Janka, N.~Saviano, K.~Scholberg, R.~Bollig, L.~Hudepohl and S.~Chakraborty,
Riv. Nuovo Cim. \textbf{39}, no.1-2, 1-112 (2016)
[arXiv:1508.00785 [astro-ph.HE]].

\bibitem{Kato:2020lwd}
C.~Kato, R.~Hirai and H.~Nagakura,
Mon. Not. Roy. Astron. Soc. \textbf{496}, no.3, 3961-3972 (2020)
[arXiv:2005.03124 [astro-ph.HE]].


\bibitem{Kato:2020hlc}
C.~Kato, K.~Ishidoshiro and T.~Yoshida,
Ann. Rev. Nucl. Part. Sci. \textbf{70}, 121-145 (2020)
[arXiv:2006.02519 [astro-ph.HE]].

\bibitem{Odrzywolek:2003vn}
A.~Odrzywolek, M.~Misiaszek and M.~Kutschera,
Astropart. Phys. \textbf{21}, 303-313 (2004)
[arXiv:astro-ph/0311012 [astro-ph]].

\bibitem{Patton:2017neq}
K.~M.~Patton, C.~Lunardini, R.~J.~Farmer and F.~X.~Timmes,
Astrophys. J. \textbf{851}, no.1, 6 (2017)
[arXiv:1709.01877 [astro-ph.HE]].

\bibitem{KamLAND:2015dbn}
K.~Asakura \textit{et al.} [KamLAND],
Astrophys. J. \textbf{818}, no.1, 91 (2016)
[arXiv:1506.01175 [astro-ph.HE]].


\bibitem{KamLAND:2024uia}
S.~Abe \textit{et al.} [KamLAND and Super-Kamiokande],
Astrophys. J. \textbf{973}, no.2, 140 (2024)
[arXiv:2404.09920 [hep-ex]].

\bibitem{SNEWS:2020tbu}
S.~Al Kharusi \textit{et al.} [SNEWS],
New J. Phys. \textbf{23}, no.3, 031201 (2021)
[arXiv:2011.00035 [astro-ph.HE]].


\bibitem{Loredo:2001rx}
T.~J.~Loredo and D.~Q.~Lamb,
Phys. Rev. D \textbf{65}, 063002 (2002)
[arXiv:astro-ph/0107260 [astro-ph]].

\bibitem{Li:2020ujl}
S.~W.~Li, L.~F.~Roberts and J.~F.~Beacom,
Phys. Rev. D \textbf{103}, no.2, 023016 (2021)
[arXiv:2008.04340 [astro-ph.HE]].

\bibitem{Bethe:1985sox}
H.~A.~Bethe and J.~R.~Wilson,
Astrophys. J. \textbf{295}, 14-23 (1985)

\bibitem{Pagliaroli:2008ur}
G.~Pagliaroli, F.~Vissani, M.~L.~Costantini and A.~Ianni,
Astropart. Phys. \textbf{31}, 163-176 (2009)
[arXiv:0810.0466 [astro-ph]].

\bibitem{Alp:2018oek}
D.~Alp, J.~Larsson, C.~Fransson, R.~Indebetouw, A.~Jerkstrand, A.~Ahola, D.~Burrows, P.~Challis, P.~Cigan and A.~Cikota, \textit{et al.}
Astrophys. J. \textbf{864}, no.2, 174 (2018)
[arXiv:1805.04526 [astro-ph.HE]].

\bibitem{Cigan:2019shp}
P.~Cigan, M.~Matsuura, H.~L.~Gomez, R.~Indebetouw, F.~Abellan, M.~Gabler, A.~Richards, D.~Alp, T.~Davis and H.~T.~Janka, \textit{et al.}
Astrophys. J. \textbf{886}, 51 (2019)
[arXiv:1910.02960 [astro-ph.HE]].

\bibitem{Page:2020gsx}
D.~Page, M.~V.~Beznogov, I.~Garibay, J.~M.~Lattimer, M.~Prakash and H.~T.~Janka,
Astrophys. J. \textbf{898}, no.2, 125 (2020)
[arXiv:2004.06078 [astro-ph.HE]].

\bibitem{XENON:2024ijk}
E.~Aprile \textit{et al.} [XENON],
Phys. Rev. Lett. \textbf{133}, no.19, 191002 (2024)
[arXiv:2408.02877 [nucl-ex]].

\bibitem{Tamborra:2014hga}
I.~Tamborra, G.~Raffelt, F.~Hanke, H.~T.~Janka and B.~Mueller,
Phys. Rev. D \textbf{90}, no.4, 045032 (2014)
[arXiv:1406.0006 [astro-ph.SR]].

\bibitem{Muller:2014rpb}
B.~M\"uller and H.~T.~Janka,
Astrophys. J. \textbf{788}, 82 (2014)
[arXiv:1402.3415 [astro-ph.SR]].

\bibitem{Walk:2019miz}
L.~Walk, I.~Tamborra, H.~T.~Janka, A.~Summa and D.~Kresse,
Phys. Rev. D \textbf{101}, no.12, 123013 (2020)
[arXiv:1910.12971 [astro-ph.HE]].

\bibitem{Mezzacappa:2020pkk}
A.~Mezzacappa,
IAU Symp. \textbf{362}, 215-227 (2020)
[arXiv:2205.13438 [astro-ph.SR]].

\bibitem{GalloRosso:2017hbp}
A.~Gallo Rosso, F.~Vissani and M.~C.~Volpe,
JCAP \textbf{11}, 036 (2017)
[arXiv:1708.00760 [hep-ph]].

\bibitem{deGouvea:2019goq}
A.~de Gouv\^ea, I.~Martinez-Soler and M.~Sen,
Phys. Rev. D \textbf{101}, no.4, 043013 (2020)
[arXiv:1910.01127 [hep-ph]].

\bibitem{Das:2017iuj}
A.~Das, A.~Dighe and M.~Sen,
JCAP \textbf{05}, 051 (2017)
[arXiv:1705.00468 [hep-ph]].

\bibitem{Mezzacappa:2024zph}
A.~Mezzacappa and M.~Zanolin,
[arXiv:2401.11635 [astro-ph.HE]].

\bibitem{Balantekin:2006tg}
A.~B.~Balantekin and Y.~Pehlivan,
J. Phys. G \textbf{34}, 47-66 (2007)
doi:10.1088/0954-3899/34/1/004
[arXiv:astro-ph/0607527 [astro-ph]].

\bibitem{Volpe:2013uxl}
C.~Volpe, D.~V\"a\"an\"anen and C.~Espinoza,
Phys. Rev. D \textbf{87}, no.11, 113010 (2013)
[arXiv:1302.2374 [hep-ph]].

\bibitem{Sawyer:2015dsa}
R.~F.~Sawyer,
Phys. Rev. Lett. \textbf{116}, no.8, 081101 (2016)
[arXiv:1509.03323 [astro-ph.HE]].

\bibitem{Capozzi:2018clo}
F.~Capozzi, B.~Dasgupta, A.~Mirizzi, M.~Sen and G.~Sigl,
Phys. Rev. Lett. \textbf{122}, no.9, 091101 (2019)
[arXiv:1808.06618 [hep-ph]].

\bibitem{Johns:2021qby}
L.~Johns,
Phys. Rev. Lett. \textbf{130}, no.19, 191001 (2023)
[arXiv:2104.11369 [hep-ph]].

\bibitem{Hyper-Kamiokande:2021frf}
K.~Abe \textit{et al.} [Hyper-Kamiokande],
Astrophys. J. \textbf{916}, no.1, 15 (2021)
[arXiv:2101.05269 [astro-ph.IM]].

\bibitem{Saez:2024ayk}
M.~M.~Saez, E.~Rrapaj, A.~Harada, S.~Nagataki and Y.~Z.~Qian,
[arXiv:2401.02531 [astro-ph.HE]].

\bibitem{Olsen:2022pkn}
J.~Olsen and Y.~Z.~Qian,
Phys. Rev. D \textbf{105}, no.8, 083017 (2022)
[arXiv:2202.09975 [astro-ph.HE]].

\bibitem{Abbar:2024nhz}
S.~Abbar and M.~C.~Volpe,
[arXiv:2401.10851 [astro-ph.HE]].

\bibitem{Hart:2023rqs}
K.~Hart, B.~J.~Shappee, D.~Hey, C.~S.~Kochanek, K.~Z.~Stanek, L.~Lim, S.~Dobbs, M.~Tucker, T.~Jayasinghe and J.~F.~Beacom, \textit{et al.}
[arXiv:2304.03791 [astro-ph.IM]].

\bibitem{Ivanez-Ballesteros:2024nws}
P.~Iv\'a\~nez-Ballesteros and M.~C.~Volpe,
[arXiv:2410.11517 [hep-ph]].

\end{thebibliography}
\end{document}